\documentclass[10pt,twocolumn,letterpaper]{article}

\usepackage{cvpr}
\usepackage{times}
\usepackage{epsfig}
\usepackage{graphicx}
\usepackage{amsmath}
\usepackage{amssymb}

\usepackage{booktabs}
\usepackage{multirow}
\usepackage{array}
\usepackage{subfig}
\usepackage{xcolor}
\usepackage{xspace}
\usepackage{bbm} 
\usepackage{appendix}

\usepackage[pagebackref=true,breaklinks=true,letterpaper=true,colorlinks,bookmarks=false]{hyperref}

 \cvprfinalcopy % *** Uncomment this line for the final submission

\def\cvprPaperID{5078} % *** Enter the CVPR Paper ID here

\ifcvprfinal\fi

\newcolumntype{C}{>{\centering\arraybackslash}p{5em}}
\newcolumntype{D}{>{\centering\arraybackslash}p{4em}}
\newcolumntype{E}{>{\centering\arraybackslash}p{4.5em}}
\newcolumntype{F}{>{\centering\arraybackslash}p{6em}}
\newcommand{\perturbation}{adversarial point perturbation\xspace}
\newcommand{\generation}{adversarial point generation\xspace}
\newcommand{\cluster}{adversarial clusters\xspace}
\newcommand{\object}{adversarial objects\xspace}

\newif\ifsubmit

\submitfalse

\ifsubmit
\newcommand{\bo}[1]{}
\newcommand{\chong}[1]{}
\newcommand{\wh}[1]{}
\newcommand{\rqi}[1]{}
\else
\newcommand{\bo}[1]{{\textcolor{blue}{[Bo: #1]}}}
\newcommand{\chong}[1]{{\textcolor{cyan}{[Chong: #1]}}}
\newcommand{\wh}[1]{{\textcolor{green}{[Warren: #1]}}}
\newcommand{\rqi}[1]{{\textcolor{red}{[Charles: #1]}}}
\fi

\begin{document}

%%%%%%%%% TITLE

\title{Generating 3D Adversarial Point Clouds}
%\title{Generating 3D Adversarial Point Clouds\\Supplementary Material}

\author{Chong Xiang \\ Shanghai Jiao Tong University \\Shanghai, China\\{\tt\small xiangchong97@gmail.com} \and Charles R. Qi\\ Facebook AI Research\\California, USA\\{\tt\small charlesq34@gmail.com} \and Bo Li\\ {\normalsize University of Illinois at Urbana-Champaign}\\Illinois, USA\\{\tt\small lxbosky@gmail.com}}
\maketitle

%%%%%%%%% ABSTRACT
\begin{abstract}

Deep neural networks are known to be vulnerable to adversarial examples which are carefully crafted instances to cause the models to make wrong predictions.
While adversarial examples for 2D images and CNNs have been extensively studied, less attention has been paid to 3D data such as point clouds.
Given many safety-critical 3D applications such as autonomous driving, it is important to study how adversarial point clouds could affect current deep 3D models.
In this work, we propose several novel algorithms to craft adversarial point clouds against PointNet, a widely used deep neural network for point cloud processing. Our algorithms work in two ways: \perturbation and \generation. For point perturbation, we shift existing points negligibly. For point generation, we generate either a set of independent and scattered points or a small number (1-3) of point clusters with meaningful shapes such as balls and airplanes which could be hidden in the human psyche.
In addition, we formulate six perturbation measurement metrics tailored to the attacks in point clouds and conduct extensive experiments to evaluate the proposed algorithms on the ModelNet40 3D shape classification dataset. Overall, our attack algorithms achieve a success rate higher than 99\% for all targeted attacks \footnote{Untargeted attacks are easier to achieve with the proposed methods, so in this paper we only focus on targeted attacks.}.
\end{abstract}

%%%%%%%%% BODY TEXT

\begin{figure}
    \centering
    \includegraphics[width =\linewidth]{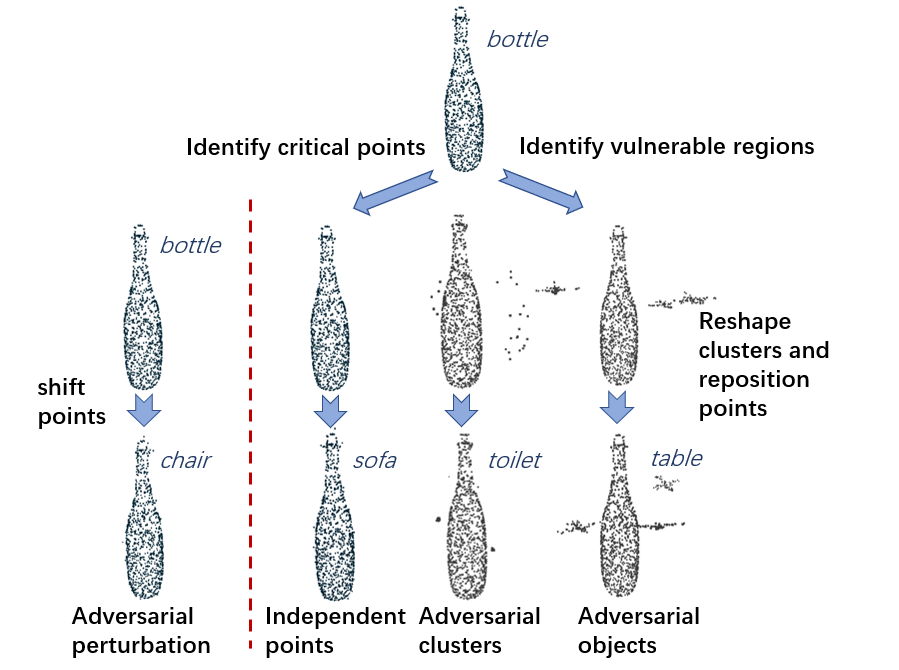}
    \caption{Attack pipeline. Our algorithms create adversarial examples by either \perturbation (left) or \generation (right). The bottle is mis-classified after our attacks.}
    \label{pipeline}
\end{figure}
\section{Introduction}

Despite of the great success in various learning tasks, deep neural networks (DNNs) have been found vulnerable to adversarial examples. The adversary is able to add imperceivable perturbation to the original data and mislead DNNs with high confidence. Many algorithms have been proposed to generate adversarial examples for data such as 2D images~\cite{szegedy2013intriguing,goodfellow2014explaining,papernot2016limitations,moosavi2016deepfool,carlini2017towards}, natural languages~\cite{jia2017adversarial,zhao2017generating}, and audios~\cite{carlini2018audio,das2018adagio}. Several recent works~\cite{athalye2017synthesizing,evtimov2017robust} have proposed adversarial examples in the 3D space, but they simply project 3D objects to 2D images as data pre-processing. One concurring work~\cite{su2018deeper} provides a simple analysis of 3D adversarial examples, but no new attack algorithm is proposed. To the best of our knowledge, we are the first to extensively study the robustness of actual 3D models which directly deal with 3D objects and propose effective attack algorithms.
Specifically, we choose to represent 3D objects with \emph{point clouds}, which are the raw data from most 3D sensors such as depth cameras and Lidars. Therefore, we attack 3D models by generating 3D adversarial point clouds.

As to the attacking target, we focus on the commonly used PointNet model~\cite{qi2016pointnet}. We choose PointNet because the model and its variants have been widely and successfully adopted in many applications such as 3D object detection for autonomous driving~\cite{qi2018frustum,zhou2018voxelnet,xu2018pointfusion}, semantic segmentation for indoor scene understanding~\cite{landrieu2017large,qi2016pointnet}, and AI-assisted shape design~\cite{sung2017complementme}. Furthermore, the model has been shown to be robust to various input point perturbations and corruptions~\cite{qi2016pointnet}. The demonstrated more robustness than 3D CNNs in~\cite{qi2016pointnet,su2018deeper} makes it a challenging and solid benchmark model for our evaluations. Although we focus on attacking PointNet, we expect our attacking algorithms and evaluation metrics extensible to more 3D models.

As the input to PointNet, a point cloud is a 3D geometric data structure that has the advantages of simple representation and low storage requirement. However, it is challenging to generate adversarial point clouds given its special properties. The point cloud's irregular format has made existing attack algorithms designed for 2D images unsuitable: 1) In raw point clouds with $XYZ$, there are no ``pixel values" positioned in a regular structure that can be slightly modified; 2) The search space for generating new adversarial points is very large, as points can be added to arbitrary positions; 3) The commonly used $L_p$ norm measurement in 2D images to bound perturbations does not fit for point cloud data with irregularity and varying cardinality.

To the best of our knowledge, we are the first to extend adversarial attack research to the irregular point cloud data, by addressing aforementioned challenges. We propose several novel attack methods for mainly two types of adversarial attacks on point clouds: \emph{\perturbation} and \emph{\generation} which are unnoticeable to human or hidden in the human psyche. The attack pipeline is illustrated in Figure~\ref{pipeline}.

For \perturbation, we propose to shift existing points negligibly. We optimize the perturbation vector under the commonly used $L_p$ norm constraint.
Our experiments show that we are able to craft unnoticeable adversarial point clouds with 100\% success rate given an acceptable perturbation budget.

For \generation, We propose to synthesize and place a set of independent points or a limited number of point clusters close to the original object. In particular, we search for ``vulnerable'' regions of objects and optimize the positions of points and the shapes of the clusters. In total, we have three kinds of generated points, namely \emph{independent points}, \emph{\cluster} (points in generic shapes such as balls), and \emph{\object} (points in object shapes such as mini airplanes), shown in the right three columns in Figure~\ref{pipeline}.
We constrain the generation by their sizes, their distances to the object surface as well as how many shapes we place.

Our attacks achieve 100\% success rate for scattered adversarial points, 99.3\% for adding three adversarial shapes, 98.2\% for two, and 78.8\% for one. 

Furthermore, in Section~\ref{sec:discussion} we discuss the transferrability of our 3D adversarial point clouds as well as the possibility to combine PointNet with CNNs to defense attacks in images. Sample code and data will be released to support further research.

To summarize, the contributions of this paper can be summarized as follows:
\begin{itemize}
\setlength\itemsep{-0.2em}

\item We are the first to generate 3D adversarial examples against 3D learning models and provide baseline evaluations for future research. Specifically, we choose representative point cloud data and PointNet model for our evaluations.
\item We demonstrate the unique challenges in dealing with irregular data structures such as point clouds and propose novel algorithms for both \emph{\perturbation} and \emph{\generation}.
\item We propose six different perturbation metrics tailored to different attack tasks and perform extensive experiments to show our attack algorithms can achieve a success rate higher than 99\% for all targeted attacks. 
\item We provide robustness analysis for 3D point cloud models and show that analyzing properties of different 3D models sheds light on potential defenses for 2D instances. 
\end{itemize}

\section{Related Work}
\label{related_work}
{\noindent \bf Point Clouds and PointNet.} Point clouds are consisted of unordered points with varying cardinality, which makes it hard to be consumed by neural networks. Qi et al.~\cite{qi2016pointnet} addressed this problem by proposing a new network called PointNet, which is now widely used for deep point cloud processing. PointNet and its variants~\cite{qi2017pointnetplusplus,wang2018dynamic} exploit a single symmetric function, \textit{max pooling}, to reduce the unordered and varying length input to a fixed-length global feature vector and thus enables end-to-end learning.
\cite{qi2016pointnet} also tried to demonstrate the robustness of the proposed PointNet and introduced the concept of critical points and upper bounds. They showed that points sets laying between critical points and upper bounds yield the same global features and thus PointNet is robust to missing points and random perturbation. However, they did not study the robustness of PointNet against adversarial manipulations, which is the main focus of this paper.

{\noindent \bf Adversarial Examples.} Szegedy et al.~\cite{szegedy2013intriguing} first pointed out that machine learning models such as neural networks were vulnerable to carefully crafted adversarial perturbation. An adversarial example which appears similar to its original data can easily fool the neural networks with high confidence. Such vulnerability of machine learning models has raised great concerns in the community and many works have been proposed to improve the attack performance~\cite{goodfellow2014explaining,papernot2016limitations,moosavi2016deepfool,carlini2017towards,papernot2017practical,xiao2018spatially,xiao2018generating} and search for possible defense~\cite{carlini2016defensive,meng2017magnet,xu2017feature,ranjan2017improving,samangouei2018defense,yan2018deepdefense}. The state-of-the-art attack algorithm, optimization based attack~\cite{carlini2017towards}, defines an objective loss function which measures both attack effectiveness and perturbation magnitude, and uses optimization to find a near-optimal adversarial solution. However, the algorithm only deals with 2D data. 
Several recent works~\cite{kurakin2016adversarial,athalye2017synthesizing,evtimov2017robust} also study the adversarial examples in the physical world. However, these works only project physical objects to 2D images and do not study models which directly deal with 3D objects.
One concurring work~\cite{su2018deeper} does provide a simple analysis of 3D adversarial examples and demonstrates PointNet's better robustness against the adversarial attacks than 3D CNNs. However, it simply adopts existing algorithms and the attack against PointNet model is not effective. To the best of our knowledge, we are the first to extensively study adversarial examples for 3D machine learning models and to propose effective algorithms for 3D adversarial point cloud generation in particular.

\section{Problem Formulation}
\paragraph{Point Cloud Data.} A point cloud is a set of points which are sampled from object surfaces. A data record $x \in \mathbb{R}^{n \times 3}$ corresponds to a point set of size $n$, where each point is represented by a 3-tuple $(x,y,z)$ coordinate.
One most important characteristic of point cloud data is its irregularity (a point cloud is not defined on a regular grid structure), which makes it hard to adapt existing attacking algorithms from 2D images.
Moreover, we are able to \emph{add points} at any positions in the 3D space while we cannot add pixels in 2D images. However, such lack of constrain results in an extremely large search space for generative adversarial examples. New attack algorithms should be proposed to address the above problems.

\paragraph{Targeted Adversarial Attacks.} In this paper, we only focus on targeted attacks against 3D point cloud classification models. It is flexible to extend our algorithms to other tasks like attacking segmentation models.

The goal of targeted attacks is to mislead a 3D deep model (e.g., PointNet) to classify an adversarial example as a selected target class.
Formally, for a classification model $\mathcal{F}:\mathcal{X}\rightarrow\mathcal{Y}$, which maps an input $x \in \mathcal{X} \subset \mathbb{R}^{n \times 3}$ to its corresponding class label $y \in \mathcal{Y} \subset \mathbb{Z}$, an adversary has a malicious target class $t^{\prime} \in \mathcal{Y}$. Based on a perturbation metric $\mathcal{D}:\mathbb{R}^{n \times 3} \times \mathbb{R}^{n^{\prime}\times 3} \rightarrow \mathbb{R}$, the goal of the attack is to find a legitimate input $x^{\prime} \subset \mathbb{R}^{n^{\prime} \times 3}$ which:
\begin{equation}%\small
        \min \mathcal{D}(x,x^{\prime}), 
        \qquad s.t. \; \mathcal{F}(x^{\prime})=t^{\prime}
\end{equation}
Note that, for point cloud data, $n$ does not necessarily equal to $n^{\prime}$. 

As mentioned in ~\cite{carlini2017towards}, directly solving this problem is difficult. Therefore, we reformulate the problem as gradient-based optimization algorithms:
\begin{equation}
\label{objective}%\small
        \min f(x^{\prime}) + \lambda*\mathcal{D}(x,x^{\prime})
\end{equation}
Here $f(x^{\prime})=(\max \limits_{i \neq t^{\prime}}(\mathcal{Z}(x^{\prime})_{i}) - \mathcal{Z}(x^{\prime})_{t^{\prime}})^+ $ is the adversarial loss function whose output measures the possibility of a successful attack, where $\mathcal{Z}(x)_i$ is the $i^{th}$ element of the logits (the input of softmax layer) and $(r)^+$ represents $\max(r,0)$. 
By optimizing over Equation~\ref{objective}, we aim to search for adversarial examples with least 3D perturbation. 

\paragraph{Attacking Types.}

In this paper, we consider two different types of attacks in point clouds~\footnote{To guarantee the points can still cover the object surface, we do not allow an adversary to remove points.}: adversarial point perturbation and adversarial point generation. In perturbation attacks, we modify existing points by shifting their $XYZ$ positions with adversarial jitters such that a point $x_i \in \mathbb{R}^3$ in the point cloud $x$ becomes ${x'}_i = x_i + \delta_i$, for $i=1,...,n$ where $\delta_i \in \mathbb{R}^3$ is the perturbation to the $i$-th point.
In generation attacks, we generate a set of adversarial points $z = \{z_i | i=1,...,k\}$ (or $z \in \mathbb{R}^{k \times 3}$ as an array representation of it) where each $z_i \in \mathbb{R}^3$ is a new point in addition to the existing point cloud $x$. Then the union of the original points and adversarial points are input to the model: $x' = x \cup z$, or in the array representation $x' \in \mathbb{R}^{(n+k) \times 3}$ through array concatenation (thus $n' = n+k$). This manner of attacking is very new and vastly different from attacks in images, because we cannot generate new pixels in a fix-sized image.

\section{Adversarial Point Perturbation}

In this section, we focus on the first and the simpler type of point cloud attack: \perturbation.
Since for perturbation we have correspondences between the original points and the perturbed ones, we can simply use $L_p$ norm to measure the distance between the two clouds.

{\noindent \bf $L_p$ Norm.} The $L_p$ norm is a commonly used metric for adversarial perturbation of fixed-shape data. For the original point sets $\mathcal{S}$ and corresponding adversarial set $\mathcal{S}^{\prime}$, the $L_p$ norm of the perturbation is defined as:
\begin{equation}%\small
\mathcal{D}_{L_p}(\mathcal{S},\mathcal{S}^{\prime})=(\sum_{i}{({s}_i-{s}_i^{\prime})}^p)^{\frac{1}{p}}
\end{equation}

%\begin{equation}
%    {\lVert v %\rVert}_p=(\sum_{i}v_i^p)^{\frac{1}{p}}
%\end{equation}
where ${s}_i$ is the $i^{th}$ point coordinate in set $\mathcal{S}$, and ${s}_i^{\prime}$ is its corresponding point in set $\mathcal{S}^{\prime}$.

% After choosing the appropriate perturbation metric $\mathcal{D}$, we are able to take advantage of
We can directly use Equation~\ref{objective} to generate the adversarial perturbations $\{\delta_i\}_{i=1}^{n}$, by optimization with the $L_2$ norm distance to bound the perturbation.

\section{Adversarial Point Generation}

Besides perturbing existing points, another general type of attacking strategy is to generate new adversarial points to mislead the 3D model. Among the ways to generate new points, a simple approach is to add arbitrary number of \emph{independent points} (Section~\ref{sec:unnotice}), ideally close to the object surface so that they are unnoticeable~\footnote{How to realize this attack in real world is still a question though.}.

On the other hand, we consider a more challenging attack task where the adversary is only able to add a limited number (1-3) of adversarial shapes (Section~\ref{sec:cluster} and Section~\ref{sec:object}), as either generic primitive shapes such as balls or meaningful shapes such as small airplane models. The task is challenging since points can only be added within small regions of the 3D space and the points of original object remain unchanged. The goal of this attack is to generate adversarial point clusters that are plausible so they cloud be hidden in the human psyche.

In the following subsections, we will introduce metrics and our attacking algorithms to generate adversarial individual points, as well as two kinds of adversarial shapes: \emph{\cluster} and \emph{\object}.
% Each subsection begins with the perturbation metrics used and followed by the point cluster generation approach.

\subsection{Generating Adversarial Independent Points}
\label{sec:unnotice}
% \chong{I change the section title}
In this section, we focus on the attack of generating (unnoticeable) independent points. Note that when adding new points to the original point clouds, we have to deal with data dimensionality changes. We first introduce metrics that measure the deviation of adversarial points to the original one and then desrcribe our attack algorithm.

\subsubsection{Perturbation Metrics}

{\noindent \bf Hausdorff Distance.} Hausdorff distance is often used to measure how far two subsets of a metric space  are from each other. Formally, for an original point set $\mathcal{S}$ and its adversarial counterpart $\mathcal{S}^{\prime}$, we define Hausdorff distance as:
\begin{equation}\label{hausdorff}%\small
    \mathcal{D}_{\textit{H}}(\mathcal{S},\mathcal{S}^{\prime})=\max\limits_{y \in \mathcal{S}^{\prime}} \min \limits_{x \in \mathcal{S}} {\lVert x-y \rVert}_2^2
\end{equation}
Intuitively, Hausdorff distance finds the nearest original point for each adversarial point and outputs the maximum square distance among all such nearest point pairs. We do not include the term $\max\limits_{x \in \mathcal{S}} \min \limits_{y \in \mathcal{S}^{\prime}} {\lVert x-y \rVert}_2^2$ since we do not modify the original object $\mathcal{S}$.

{\noindent \bf Chamfer Measurement.\footnote{We name it as ``Chamfer measurement" since this perturbation metric does not satisfy triangle inequality, which means it does not satisfy the definition of distance.}} Chamfer measurement~\cite{fan2017point} is a similar perturbation metric as Hausdorff distance. The difference is Chamfer Measurement takes the average, rather than the maximum, of the distances of all nearest point pairs. The formal definition is as follows:
\begin{equation}\label{chamfer}%\small
    \mathcal{D}_{\textit{C}}(\mathcal{S},\mathcal{S}^{\prime})=\frac{1}{{\lVert \mathcal{S}^{\prime} \rVert}_0} \sum\limits_{y \in \mathcal{S}^{\prime}} \min \limits_{x \in \mathcal{S}} {\lVert x-y \rVert}_2^2
\end{equation}

{\noindent \bf Number of Points Added.} We also want to measure the number of points added in our attack, by counting points whose distances from the object surface is above a certain threshold. Formally, for an original point set $\mathcal{S}$, the generated point set $\mathcal{S}^{\prime}$, and a threshold value $T_\textit{thre}$, the number of points added is defined as:
\begin{equation}
\label{count}%\small
    \textit{Count}(\mathcal{S},\mathcal{S}^{\prime})= \sum_{y \in \mathcal{S}^{\prime}}\mathbbm{1}[ \min\limits_{x \in \mathcal{S}}{\lVert x-y \rVert}_2 > T_\textit{thre}]
\end{equation}
where $\mathbbm{1}[\cdot]$ is the indicator function whose value is 1 when the statement is true and 0 otherwise. Note that the number of points added is not optimized as the perturbation metric $\mathcal{D}$ in Equation~\ref{objective} due to its incompatibility with gradient-based optimization algorithms, but is reported as an additional performance metric. %Only Hausdorff and Chamfer distance are implemented as the perturbation metric $\mathcal{D}$ in Equation~\ref{objective}. 

\subsubsection{Attacking Algorithm}

Directly adding points to the unconstrained 3D space is infeasible due to the large search space. Therefore we propose an \textit{initialize-and-shift} method to find appropriate position for each added point:
\begin{enumerate}
\setlength\itemsep{-0.2em}
    \item \textit{Initialize} a number of points to the same coordinates of existing points as initial points.
    \item \textit{Shift} initial points via optimizing Equation~\ref{objective} and output their final positions.
\end{enumerate}

During the optimization process, some initial points are shifted from their initial positions and ``added" to the original objects as adversarial points. 
The others that are barely shifted do not change the shape of the object, and thus can be discarded as points-not-added.

To make the optimization more efficient, we propose to initialize points to the positions of ``critical points'' of the target. Critical points are like key points or salient points in a 3D point cloud. In PointNet specifically, they can be computed by taking the points that remain active after the max pooling~\cite{qi2016pointnet}, which means they are at important positions that determine the object category. Adversarial points around these critical positions are more likely to change the final prediction.

We use Hausdorff and Chamfer measurements as the perturbation metrics $\mathcal{D}$ for this attack because they are more capable of measuring how unnoticeable the adversarial point clouds of different dimensionality are.

\subsection{Generating Adversarial Clusters}
\label{sec:cluster}

For \emph{\cluster}, we aim to minimize the radius of the generated cluster so that they look like a ball attached to the original object and will not arouse suspicion. In addition, we also encourage the cluster to be close to the object surface. To satisfy these two requirements, we introduce the perturbation metrics used as follows.

\subsubsection{Perturbation Metrics}

{\noindent \bf Farthest Distance.} If the farthest pair-wise point distance in a point set is controlled within a certain threshold, the points in this set are able to form a shaped cluster. Formally, we define farthest distance of a point set $\mathcal{S}$ as:
\begin{equation}%\small
    \mathcal{D}_{\textit{far}}(\mathcal{S})= \max\limits_{x,y \in \mathcal{S}} {\lVert x-y \rVert}_2
\end{equation}

{\noindent \bf Chamfer Measurement.} Besides encouraging point cluster to form within a small radius, we may also want to push the added clusters towards the surface of the object. Therefore, we also include the Chamfer Measurement (defined in Equation~\ref{chamfer}) as our perturbation metric and optimization objective.

{\noindent \bf Number of Clusters Added.} Similar to the number of points added in the unnoticeable adversarial point cloud generation, the number of clusters added also serves as an additional metric for attack performance, which is hard bounded to 1-3 in our experiments.

\subsubsection{Attacking Algorithm}
Before going into the details of generation algorithms, we need to reformulate Equation~\ref{objective} as follow:
\begin{equation}
\label{obj_revised}%\small
\min f(x^{\prime}) + \lambda\cdot ( \sum_{i}\mathcal{D}_\textit{far}(\mathcal{S}_i)+\mu\cdot\mathcal{D}_\textit{C}(\mathcal{S}_0,\mathcal{S}_i) )
\end{equation}
where $i \in \{1,2,\ldots,m\}$, $\mathcal{S}_0$ is the original object, $\mathcal{S}_i$ is the $i^{th}$ adversarial point cluster, $m$ is number of adversarial clusters, and $\mu$ is the weight used to balance the importance between Farthest Distance loss and Chamfer Measurement loss. Here we abuse the notation a little to use $\mathcal{D}$ to denote both mappings $\mathbb{R}^{n \times 3} \times \mathbb{R}^{n^{\prime}\times 3} \rightarrow \mathbb{R}$ and $\mathbb{R}^{n \times 3} \rightarrow \mathbb{R}$.

Generating adversarial clusters is a special case of adding adversarial point clouds, so we can adopt the \textit{initialize-and-shift} method used. However, unlike independent point generation, we have to constrain the added points clustered to be within small regions. As points are likely to get stuck in their initialized vicinity due to the ubiquity of local-minima, we need a more efficient initialization methods for \cluster generation.

We try to leverage the idea of ``vulnerable regions'' for initialization. For formatted data like 2D images, it is common to impose a $L_1$ constraint to encourage the sparsity of the perturbation vector. The region with large perturbation under a proper $L_1$ constraint is believed to be important for model decisions and thus vulnerable to adversarial attacks. However, the $L_1$ constraint is not well defined on point clouds thus inapplicable here. Instead, we take advantage of ``critical points'' again to effectively find potentially vulnerable regions for initialization. Critical points, as a subset of the original set, collectively determine the global features of the object shape but could also be vulnerable regions to attacks.

Given a victim object and a target class $t^{\prime}$, the attack process is as follows:
\begin{enumerate}
\setlength\itemsep{-0.2em}%\small
    \item Obtain the critical points of the objects in \emph{target} class. 
    \item Use the clustering algorithm DBSCAN~\cite{ester1996density} to cluster the selected critical points.
    \item Choose points in the $k$ largest clusters as the initial points, where $k$ is a self-chosen parameter as well as a metric for attack performance evaluation.
    \item Optimize over Equation~\ref{obj_revised} using gradient-based algorithms and find optimal cluster positions and shapes.
\end{enumerate}
Note that DBSCAN groups points that are closely packed (or points with local density passing a threshold), while marking the other points lying in low-density regions as outliers~\cite{ester1996density}. 
Thus, we are able to filter out outlier points and get compact clusters via it.

Besides tuning DBSCAN, it is essential to determine the number of objects in target class we use, as well as the number of critical points selected from each target object. Choosing only one target object restricts the space distributions of the critical points. However, using too many target objects result in density scattered critical points unhelpful to identify a sparse set of vulnerable regions. The reasons for tuning the number of critical points selected are similar, we want a moderate number of them. However, such parameter tuning does not need to too fine grained as the attack pipeline is still dominated by the optimization over Equation~\ref{obj_revised}. %We will provide detailed experimental setting in the next section.

\subsection{Generating Adversarial Objects}
\label{sec:object}
For this attack, we start from some meaningful objects like small airplanes, slightly modify them, and place them in the appropriate adversarial positions. People may not become suspicious because the adversarial objects are like other benign objects nearby.

\subsubsection{Perturbation Metrics}

{\noindent \bf $L_p$ Norm.} Since we want to only slightly modify the meaningful objects and make the generated shapes similar to the real-world ones, we adopt the $L_p$, specifically $L_2$, as our first metric.

{\noindent \bf Chamfer Measurement.} Similar to the adversarial clusters, we want to encourage the generate shape to be close the original object. 

{\noindent \bf Number of Clusters Added.} Number of clusters added, bounded to 1-3, is also used to evaluate the attack performance.

% \subsubsection{Generating Adversarial Objects}
\subsubsection{Attacking Algorithm}
We also need to rewrite the objective function to fit the attack setting:
\begin{equation}
\label{obj_revised2}%\small
\min f(x^{\prime}) + \lambda\cdot ( \sum_{i}\mathcal{D}_{L_2}(\mathcal{S}_{i0},\mathcal{S}_i)+\mu\cdot\mathcal{D}_\textit{C}(\mathcal{S}_0,\mathcal{S}_i) )
\end{equation}
where $i \in \{1,2,\ldots,m\}$, $\mathcal{S}_0$ is the original object, $\mathcal{S}_i$ is the $i^{th}$ adversarial point cluster, $\mathcal{S}_{i0}$ is the $i^{th}$ real-world clusters, $m$ is number of adversarial clusters, and $\mu$ is the weight used to balance the importance between $L_2$ loss and Hausdorff Distance loss. 
To mount this attack, we need to find the vulnerable regions first and then to initialize the perturb the added real-world point clusters. The attack pipeline is as follows:
\begin{enumerate}%
%\small
\setlength\itemsep{-0.2em}
    \item Obtain the critical points of the objects in \emph{target} class. 
    \item Use the clustering algorithm DBSCAN~\cite{ester1996density} to cluster the selected critical points.
    \item Identify the $k$ largest clusters and calculate the position of cluster centers, where $k$ is a self-chosen parameter as well as a metric for attack performance evaluation.
    \item Choose meaningful objects and initialize them to make their centers overlap with the calculated positions in the previous step.
    \item Optimize over Equation~\ref{obj_revised2} using gradient-based algorithms and find optimal cluster positions and shapes.
\end{enumerate}

Note that we also have the freedom to choose the orientation of the modified clusters. Since adversarial clusters with different orientations would not arouse suspicion, we do not impose an constraint on the magnitude of rotation.

\section{Experiment Results}
In this section, we implement the proposed algorithms for different attack tasks and conduct extensively evaluation on  attack performance based on various metrics. %Additionally, we explore the transferability of generated adversarial point clouds across different 3D learning models.

\subsection{Dataset and 3D Models} We use the aligned benchmark ModelNet40~\cite{wu20153d,SZB17a} dataset for our experiments. The ModelNet40 dataset contains 12,311 CAD models from 40 most common object categories in the world. 9,843 objects are used for training and the other 2,468 for testing. As done by Qi et al.~\cite{qi2016pointnet}, we uniformly sample 1,024 points from the surface of each object, and re-scale them into a unit ball. We use the same PointNet structure as proposed in~\cite{qi2016pointnet} and train the model with all ModelNet40 training data to obtain our victim model. The ModelNet40 dataset is very imbalanced. For our attacks, we randomly select 25 test examples from each 10 largest classes, namely airplane, bed, bookshelf, bottle, chair, monitor, sofa, table, toilet and vase, to generated adversarial point clouds for. For each victim data record, we generate adversarial examples targeted on the rest 9 classes. Therefore, we have 2,250 (victim,target) attach pairs for our experiments.

\begin{table*}[t!]
\centering
\resizebox{\textwidth}{!}{
\begin{tabular}{c | C C | C F C | C F C }
  \toprule
  
 \multirow{2}{*}{Case}  & \multicolumn{2}{c}{Shifting Points ($L_2$ Norm)}& \multicolumn{3}{c}{Adding Points ($\mathcal{D}_\textit{H}$)}& \multicolumn{3}{c}{Adding Points ($\mathcal{D}_\textit{C}$)}\\

 & mean loss & success rate & mean loss & \#points added &success rate & mean loss &\#points added& success rate\\
  \midrule
Best& 0.0874& 100\% &0.0003&93&100\%&$3.1\times{10}^{-5}$&58&100\%\\
Average&0.3032& 100\% &0.0105&88&100\%&$2.7\times{10}^{-4}$&51&100\%\\
Worst& 0.4674& 100\% &0.0210&99&100\%&$7.2\times{10}^{-4}$&49&100\%\\
  \bottomrule
\end{tabular}
}
\caption{Attack performance evaluation for \perturbation and adversarial independent point generation}
\label{tab_unnoticeable}
\end{table*}

\begin{table*}[!tbhp]
\centering
\resizebox{\textwidth}{!}{
\begin{tabular}{c| C C C | C C C | C C C }
  \toprule
  
    \multirow{2}{*}{Attack}& \multicolumn{3}{c}{\#Shape 1}& \multicolumn{3}{c}{\#Shape 2}& \multicolumn{3}{c}{\#Shape 3 }\\

&$\mathcal{D}_{\textit{far}}$ / $\mathcal{D}_{L_2}$ & $\mathcal{D}_{C}$& success rate & $\mathcal{D}_{\textit{far}}$ / $\mathcal{D}_{L_2}$& $\mathcal{D}_{C}$ & success rate & $\mathcal{D}_{\textit{far}}$ / $\mathcal{D}_{L_2}$ & $\mathcal{D}_{C}$ & success rate\\
  \midrule
  %Best& NA&NA&NA&NA&NA&NA&NA&NA&NA\%\\
 \emph{\cluster}& 0.5401&0.1374&78.8\%&0.3118&0.1839&98.2\%&0.1818&0.1744&99.3\%\\
 \emph{\object}& 0.5539&0.1776& 54.6\% &0.0838&0.1332&93.8\%&0.0212&0.0855&97.3\%\\
%Worst& NA&NA&NA&NA&NA&NA&NA&NA&NA\%\\
  \bottomrule
\end{tabular}
}
\caption{Attack performance evaluation for \cluster and \object (average case).}
\label{tab_meaingful1}
\end{table*}

\subsection{Adversarial Point Perturbation Evaluation}
\begin{figure}[!t]
    \centering
    \resizebox{\linewidth}{!}{
    \includegraphics{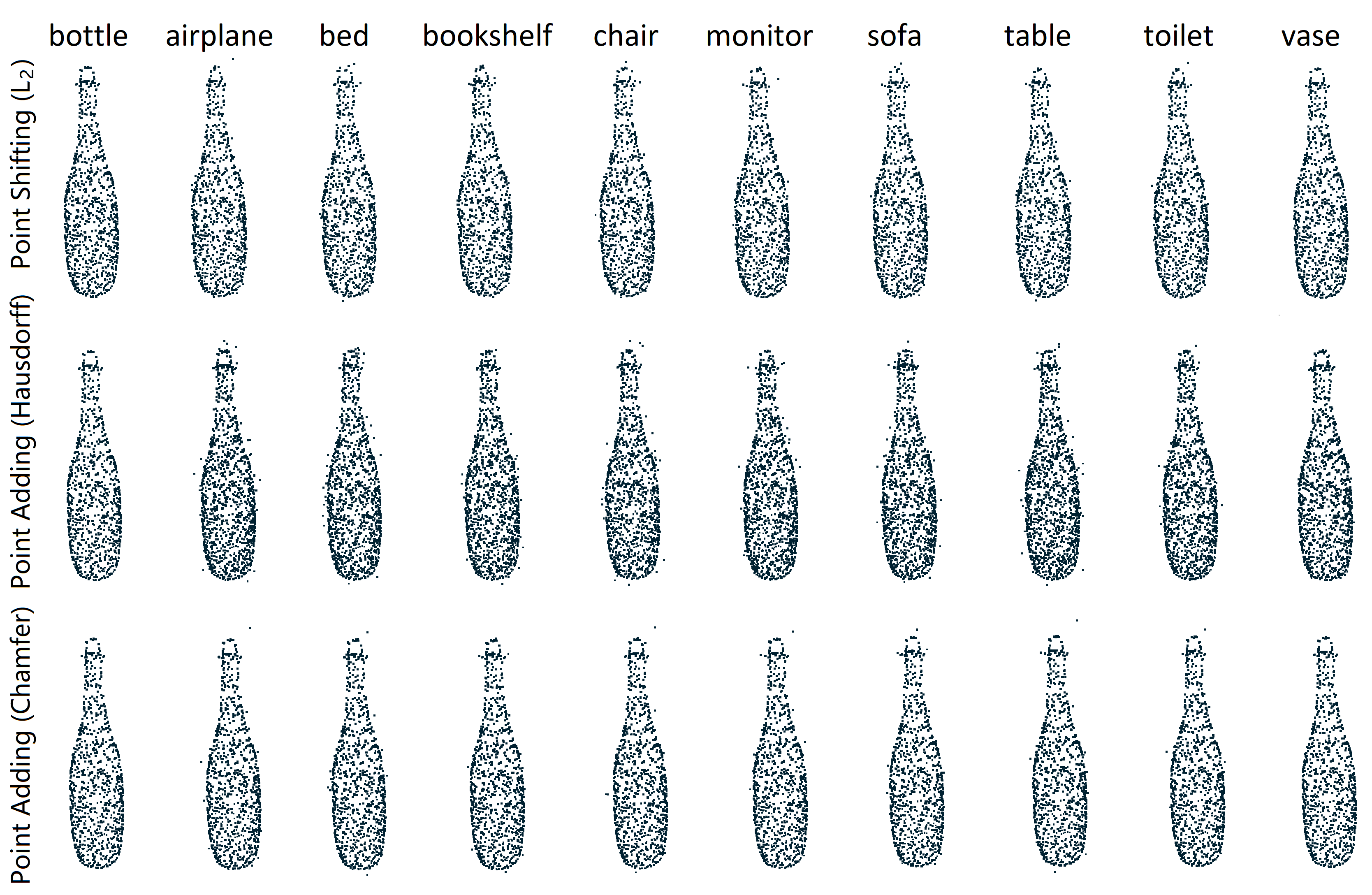}}
    \caption{Visualization for \perturbation.}
    \label{visu_unnoticeable}

\end{figure}

We evaluate the attack performance for \perturbation in this subsection.
We use $L_2$ distance as the perturbation constraint $\mathcal{D}$ in Equation~\ref{objective} and minimize the objective loss to find the optimal perturbation. To obtain good attack performance, it is essential to choose an appropriate value for the weight $\lambda$, which controls the balance between minimizing adversarial loss and perturbation magnitude. If the $\lambda$ is too small, the perturbation constraint is not strong enough and the perturbation would become too obvious. On the other hand, a $\lambda$ that is too large would result in minimizing perturbation magnitude only and fail to attack. For all of our attacks, we perform 10-step binary search for the near-optimal $\lambda$. During the search, we record the smallest perturbation $\mathcal{D}(x,x^{\prime})$ and its corresponding adversarial example $x^{\prime}$, and finally output the most unnoticeable adversarial example. 

we report the experiment results for three cases: \textit{best case} for the most easily attacked (victim,target) class pair, \textit{average case} for all attacking class pairs, and \textit{worst case} for the most difficult pair. The success rate and mean perturbation loss for point shifting attacks are reported in the first two columns of Table~\ref{tab_unnoticeable}. We can see we successfully attack all victims examples into all target classes. The perturbation loss for this attack is relatively small, considering the perturbation vector contains 1,024 elements. 
We also provide visualization in the first row of Figure~\ref{visu_unnoticeable}. We choose class ``bottle" as our visualization victim because adversarial perturbation would become more obvious for a simple shape like a bottle. More visualization for other objects can be found in the supplementary. From the visualization, we can see the perturbation (the adversarial point cloud) is nearly indistinguishable.

\begin{figure*}[!t]
    \centering
    \resizebox{\linewidth}{!}{
    \includegraphics{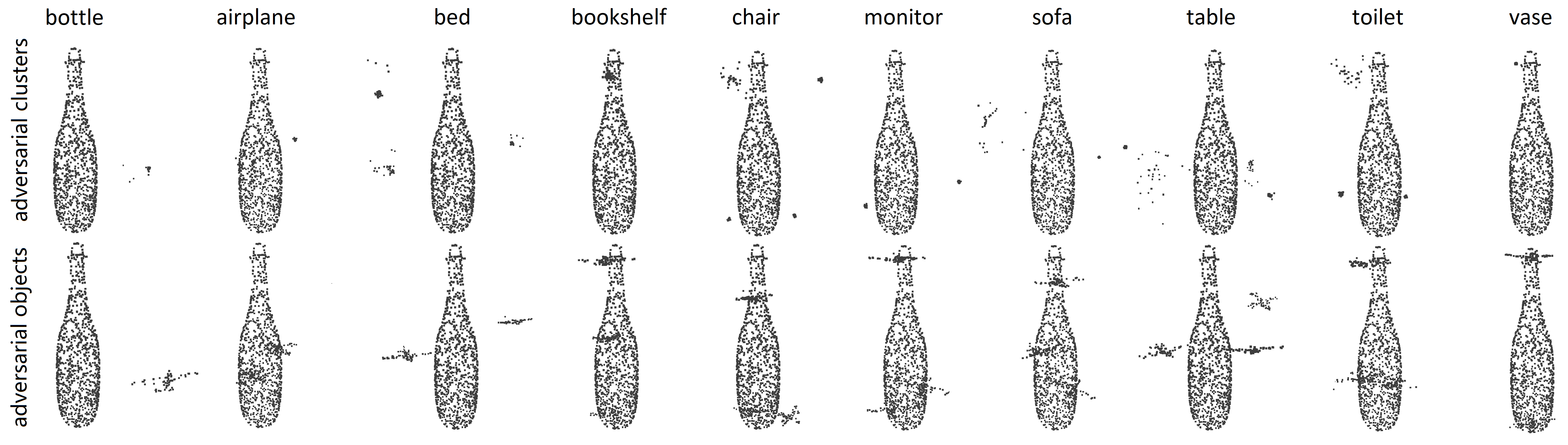}}
    \caption{Visualization for adding 3 adversarial clusters/objects.}
    \label{visu_manufacturable}

\end{figure*}

\subsection{Adversarial Point Generation Evaluation}
In this subsection, we evaluate the attack performance of three different ways of \generation: \emph{independent points}, \emph{\cluster}, and \emph{\object}.

{\noindent \bf Adversarial Independent Points.} 
For generating adversarial independent points, we take Hausdorff and Chamfer measurements as perturbation metrics $\mathcal{D}$ and optimize over Equation~\ref{objective}. We use two different distances separately and compare performance of these two constraints. To calculate the number of points added, we get the critical points of the newly generated adversarial point clouds, set $T_\textit{thre}$ to 0.01, and count over points moved further than it. The experiment results are shown in Table~\ref{tab_unnoticeable}. First, both Hausdorff and Chamfer constraints result in attack success rate of 100\%, proving the effectiveness of proposed \textit{initialize-and-shift} algorithm. Secondly, we can see great difference between the mean distance losses and number of points added. Since Hausdorff distance only controls the largest distance of all nearest point pairs while Chamfer measurement calculates the average distance, the loss value of Hausdorff is much larger than that of Chamfer and so does the number of points added. A more detailed analysis on the different attack performance of Hausdorff and Chamfer measurement is included in the supplementary.

%Another interesting finding is that the best case for Hausdorff constraint in terms of perturbation loss has more points added than the average case. This means Hausdorff distance does not explicitly constrain the number of points added.
The visualization of the adversarial bottle by adding points is shown in the second and third rows of Figure~\ref{visu_unnoticeable}. From the figure, we can easily observe the different characteristics of different constraints (Hausdorff constraint results in more added points while Chamfer constraint leads to more obvious outliers). Considering the different properties of the two constraints, one can combine these two constraints and adjust the weights for the two perturbation metrics according to the specific attack goals.

{\noindent \bf Adversarial Clusters.} To get initial clusters, we randomly select 8 different objects from the target class test set, and obtain 32 most important critical points for each selected object based on the number of global feature channels it contributes to. We then use DBSCAN algorithm to cluster these $32 \times 8$ critical points. After that, we retain $k$ clusters of largest size and discard other small clusters and outliers. For each cluster, we conduct subsampling or padding to obtain 32 initial points. We vary $k$ from $1$ to $3$ to see how the number of adversarial clusters affects the attack performance (success rate and distance loss).  In Equation~\ref{obj_revised}, the parameter $\lambda$ is chosen via 5-step binary search while $\mu$ is prefixed according to the adversary's preference on smaller or closer clusters. In our experiment, we set $\mu$ to 0.1. Due to the lack of space, we only report the quantitative results for \textit{average case} in Table~\ref{tab_meaingful1}. The comprehensive results for three cases are included in the supplementary.

The table shows that, as the number of adversarial clusters increases, the attack success rate is significantly improved and we are able to attack 99.3\% examples when adding 3 adversarial clusters. Moreover, a larger number of added clusters also helps reduce the perturbation loss for each cluster. When we only add one cluster, the farthest distance of the cluster for average case is 0.5401, which is quite large considering the whole object fits in a unit ball. However, the farthest distance drops dramatically to 0.1818 when we are adding 3 clusters. Thus, it is reasonable to expect better attack performance if the adversary is able to add more than 3 clusters. 
Visualization for adding 3 adversarial clusters can be found in the first row of Figure~\ref{visu_manufacturable}. Several small clusters are clearly shown for most attack pairs.
% \chong{however, we have to admit that for some attacking pair, it is difficult to generate small clusters. We ascribe this to the }

{\noindent \bf Adversarial Objects.} For this attack, we use the same setting as the adversarial cluster to identify the vulnerable regions. We randomly select an object from the ``airplane" class and initialize it to the centers of different vulnerable regions. The airplane is re-scaled to three-tenths of its original size, and 64 points are uniformly sampled from the surface.~\footnote{We choose an airplane to simulate the scenario where the adversary could manipulate several micro-UAVs to suspend around the victim object.} After the initialization, we optimize according to Equation~\ref{obj_revised2} to perform the attack. Similarly, the parameter $\lambda$ is determined by binary search while the $\mu$ is pre-set to $0.2$ in our experiment.

The results for different perturbation metrics are provided in the second row of Table~\ref{tab_meaingful1}. We can this attack is more challenging than that of adversarial clusters since the shapes of the objects are almost predefined. However, the shapes similar to real-world objects made this attack less suspicious. Moreover, we still achieve a reasonably high success rate of 97.3\% when adding three adversarial clusters, and we can achieve better performance if more adversarial objects are allowed to be added. 
We also provide visualization in the second row of Figure~\ref{visu_manufacturable}. We can see the several small airplanes near the bottle are already capable to fool the PointNet model. Comparison between two rows of visualization in Figure~\ref{visu_manufacturable} shows that when the clusters are close to the object, \emph{\cluster} have better visualization performance while \emph{\object} are less suspicious when the clusters are far from the surface. This further justifies our attempt to introduce two kinds of meaningful shapes.

\begin{table}[t]
\centering
\resizebox{\linewidth}{!}{
\begin{tabular}{c |EEEEE }
  \toprule
   & Shift & Add ($D_\textit{H}$) & Add ($D_\textit{C}$)  & 3 Clusters & 3 Objects \\
  \midrule
PointNet++~\cite{qi2017pointnetplusplus} & 3.9\% & 8.9\% & 3.6\% & 8.9\% &9.6\%\\
DGCNN~\cite{wang2018dynamic}    & 1.9\% & 6.8\% & 7.4\% & 16.9\% &16.5\%\\
Augmented PointNet     & 3.0\% & 4.0\% & 5.7\% & 46.5\% &34.5\%\\
Different initialization & 5.5\% & 7.4\% & 7.8\% & 48.3\% &39.2\%\\
  \bottomrule
\end{tabular}
}
\caption{Attack success rates of untargeted transfer attacks against PointNet++, DGCNN, augmented PointNet, and PointNet with a different weight initialization.}
\label{transfer}
\end{table}

\begin{table}[]
    \centering
    
    \resizebox{0.9\linewidth}{!}{\begin{tabular}{c|cccccc}
    \toprule
    $\epsilon$   &0 &0.1 & 0.2&  0.3 & 0.4 & 0.5\\
    \midrule
    LeNet (original) & 99.2\% & 70.7\% & 35.5\% & 18.7\% & 12.0\% & 9.0\% \\
    LeNet (binarized) &98.9\%&97.2\%&93.3\%& 86.2\% & 75.4\% & 28.0\%\\  
    PointNet& 99.0\% &98.6\% & 98.1\%&96.8\% & 94.1\% & 63.8\% \\
    \bottomrule
    \end{tabular}}
    \caption{Test accuracy of adversarial examples on MNIST.}
    \label{tab_mnist}
\end{table}
\subsection{More Analysis on 3D Model Robustness}
\label{sec:discussion}
In this subsection, we provide robustness analyses for PointNet-like models.

{\noindent \bf Transferablity of Adversarial Point Clouds.} %Considering the transferability of 2D adversarial instances, here we aim to explore the transferability of the generated adversarial point clouds against other PointNet-like models.
We feed our crafted adversarial point clouds to PointNet++~\cite{qi2017pointnetplusplus}, DGCNN~\cite{wang2018dynamic}, a PointNet trained with data augmentation, and a PointNet trained with a different weight initialization and find that theses 3D adversarial examples actually hardly transfer as targeted attacks. Furthermore, we calculate the success rate for untargeted attack and the results are shown Table~\ref{transfer}. We can see the transferability for untargeted attack is also limited compared with 2D adversarial examples. 
Since our proposed attack methods are general and can be applied to attack other 3D models, the low transferability may be related with special properties of 3D models themselves. This intrinsic property makes it possible to design black-box defense against such adversarial instances.

%{\noindent \bf Training with Data Augmentation.} Instead of using the original aligned training data, we randomly rotate and jitter the point clouds during the training. The untargeted attack success rate is reported in last row of Table~\ref{transfer}. The results show that the model trained with data augmentation is significantly more robust against adversarial examples.

{\noindent \bf Defense on MNIST~\cite{lecun1998gradient} with PointNet.} Motivated by aforementioned robustness analysis, we take a step forward to use PointNet structure for defense on MNIST dataset. We binarize the grey-scale images and sample 256 points from each MNIST digit. We craft adversarial examples by attacking LeNet~\cite{lecun1998gradient} with FGSM~\cite{goodfellow2014explaining}. The test accuracy of binarized images (LeNet) and sampled point clouds (PointNet) with different values of attack parameter $\epsilon$ are reported in Table~\ref{tab_mnist}. We can see the PointNet model achieve relatively high test accuracy and show promising defense properties against adversarial examples.

\textit{In summary, our analysis shows that PointNet structure is more robust than traditional CNNs. We believe part of the robustness comes from the learning of global features via max pooling. Though the intriguing properties is not fully understood yet, we believe a further study on this would motivate a defense direction to include PointNet-like structure to improve the robustness of traditional nerual networks.}

\section{Conclusion}

Arguable as the first to study the vulnerability of 3D learning models, in this paper, we have proposed several attacking algorithms to generate adversarial point clouds to fool the widely used PointNet model, including \perturbation and \generation . We also propose six different perturbation metrics and extensively evaluate the performance of the proposed attack algorithms. Our extensive experiment results show that the proposed algorithms are able to find 3D adversarial point clouds with an attack success rate higher than 99\% given an acceptable perturbation budget. We hope this work is able to provide a baseline as well as a guideline for future 3D adversarial example research.
%\newpage

 {\small
 \bibliographystyle{ieee}
 \bibliography{3d_adv}
 }

\appendices

\section{Additional Quantitative Results}
In this section, we provide additional quantitative results for point clouds attacks from \cluster, \object, \perturbation, and independent points adding.

{\noindent \bf Adversarial Clusters and Adversarial Object under Three Attack Cases.}
% The additional experiment results for \cluster and \object under three different cases are tabulated in Table~\ref{tab_meaingful1_three} and Table~\ref{tab_meaingful2_three}.
Table~\ref{tab_meaingful1_three} and Table~\ref{tab_meaingful2_three} report our attacks under three cases: best case, average case and worst case for adversarial cluster attack and adversarial object attack respectively. We report (victim,target) pairs with the least distance losses among all 100\% successfully attacked pairs as the best cases. We report the (victim, target) pairs with the smallest success rates as the worst cases. It is obvious that constraining the attack to only one cluster significantly increases the attack difficulty.

\begin{figure}
    \centering
    \includegraphics[width=0.8\linewidth]{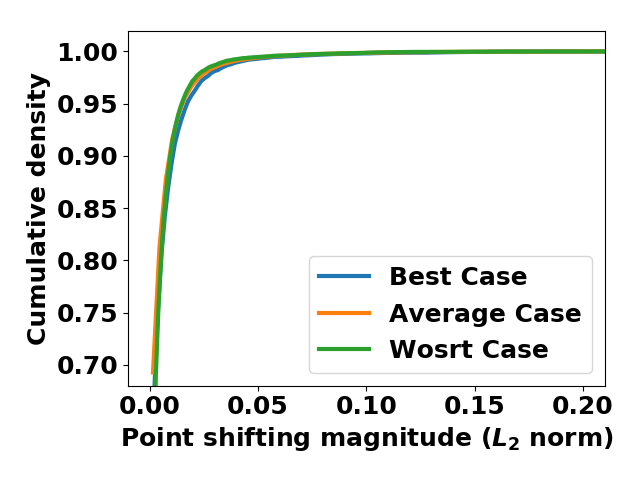}
    \caption{CDF of point shifting distance for \perturbation .}
    \label{shift_distribution}
\end{figure}

{\noindent \bf Adversarial Point Perturbation.} To better understand the attack performance of point shifting in \perturbation, we plot the distribution of perturbation magnitude ($L_2$ norm) for each point in Figure~\ref{shift_distribution}. It is obvious that for all three cases, most points (80\%) are barely shifted (less than 0.005 compared to the object scale of 1.0), and the shifting distances for most shifted point are within 0.03, which is negligible comparing with the size of a unit ball.

\begin{figure}
    \centering
    \includegraphics[width=0.8\linewidth]{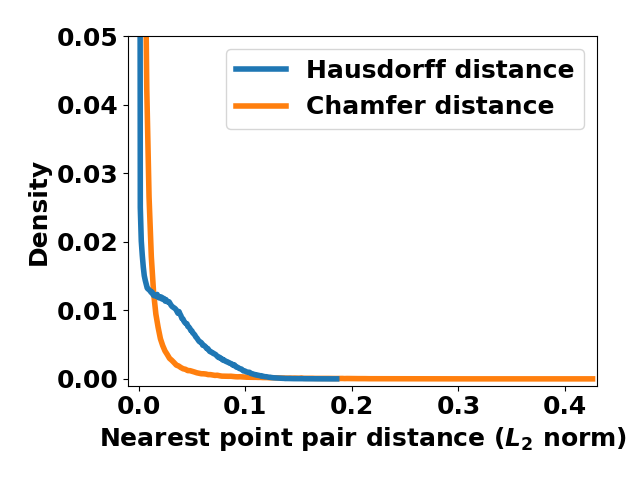}
    \caption{Distributions for distance of nearest point pairs in independent point adding attack.}
    \label{add_distribution}
\end{figure}

{\noindent \bf Adversarial Independent Points.} To help further understand the characteristics of Hausdorff and Chamfer constraints and explain why we include Hausdorff distance despite of its ``poor" quantitative performance, we plot the distribution of distances from each point to the object surface in Figure~\ref{add_distribution}. As expected, the number or percentage of points with non-trivial distance under Hausdorff optimization is larger than that under Chamfer optimization. However, it should be noticed that the largest distance of the Hausdorff case ($0.18$) is much smaller than that of Chamfer ($0.42$). This difference suggests that added points with Hausdorff constraint are likely to have fewer outliers, and thus less noticeable compared with those added based on Chamfer constraint. This result justifies our proposal to include Hausdorff distance as a perturbation metric $\mathcal{D}$.

\begin{table*}[h!]
\centering
\resizebox{\textwidth}{!}{
\begin{tabular}{c| C C C | C C C | C C C }
  \toprule
  
    \multirow{2}{*}{Case}& \multicolumn{3}{c}{1 cluster}& \multicolumn{3}{c}{2 clusters}& \multicolumn{3}{c}{3 clusters }\\

&$\mathcal{D}_{\textit{far}}$ & $\mathcal{D}_{C}$ & success rate & $\mathcal{D}_{\textit{far}}$ & $\mathcal{D}_{C}$ & success rate & $\mathcal{D}_{\textit{far}}$ & $\mathcal{D}_{C}$ & success rate\\
  \midrule
  Best& 0.0207&0.0300&100\%&0.0184&0.0331&100\%&0.0191&0.0349&100\%\\
 Average& 0.5401&0.1372&78.8\%&0.3118&0.1839&98.2\%&0.1818&0.1744&99.3\%\\
Worst& 0.0265&0.0051&4.0\%&0.4452&0.0286&64.0\%&0.4797&0.1410&80.0\%\\
  \bottomrule
\end{tabular}
}
\caption{Attack performance evaluation for \cluster (three cases).}
\label{tab_meaingful1_three}
\end{table*}

\begin{table*}[h!]
\centering
\resizebox{\textwidth}{!}{
\begin{tabular}{c| C C C | C C C | C C C }
  \toprule
    \multirow{2}{*}{Case}& \multicolumn{3}{c}{1 object}& \multicolumn{3}{c}{2 objects}& \multicolumn{3}{c}{3 objects}\\

&$\mathcal{D}_{L_2}$ & $\mathcal{D}_{C}$ & success rate & $\mathcal{D}_{L_2}$ & $\mathcal{D}_{C}$ & success rate & $\mathcal{D}_{L_2}$ & $\mathcal{D}_{C}$ & success rate\\
  \midrule
  Best& 0.0071&0.0176&100\%&0.0019&0.0072&100\%&0.0021&0.0073&100\%\\
 Average& 0.5539&0.1776&54.6\%&0.0838&0.1332&93.8\%&0.0212&0.0850&97.3\%\\
Worst& 0.1256&0.0223&8.0\%&0.0883&0.0205&20.0\%&0.0485&0.0832&56.0\%\\
  \bottomrule
\end{tabular}
}
\caption{Attack performance evaluation for \object (three cases).}
\label{tab_meaingful2_three}
\end{table*}

\section{Additional Visualization Results}
\begin{figure*}
    \centering
    \includegraphics[width=0.8\textwidth]{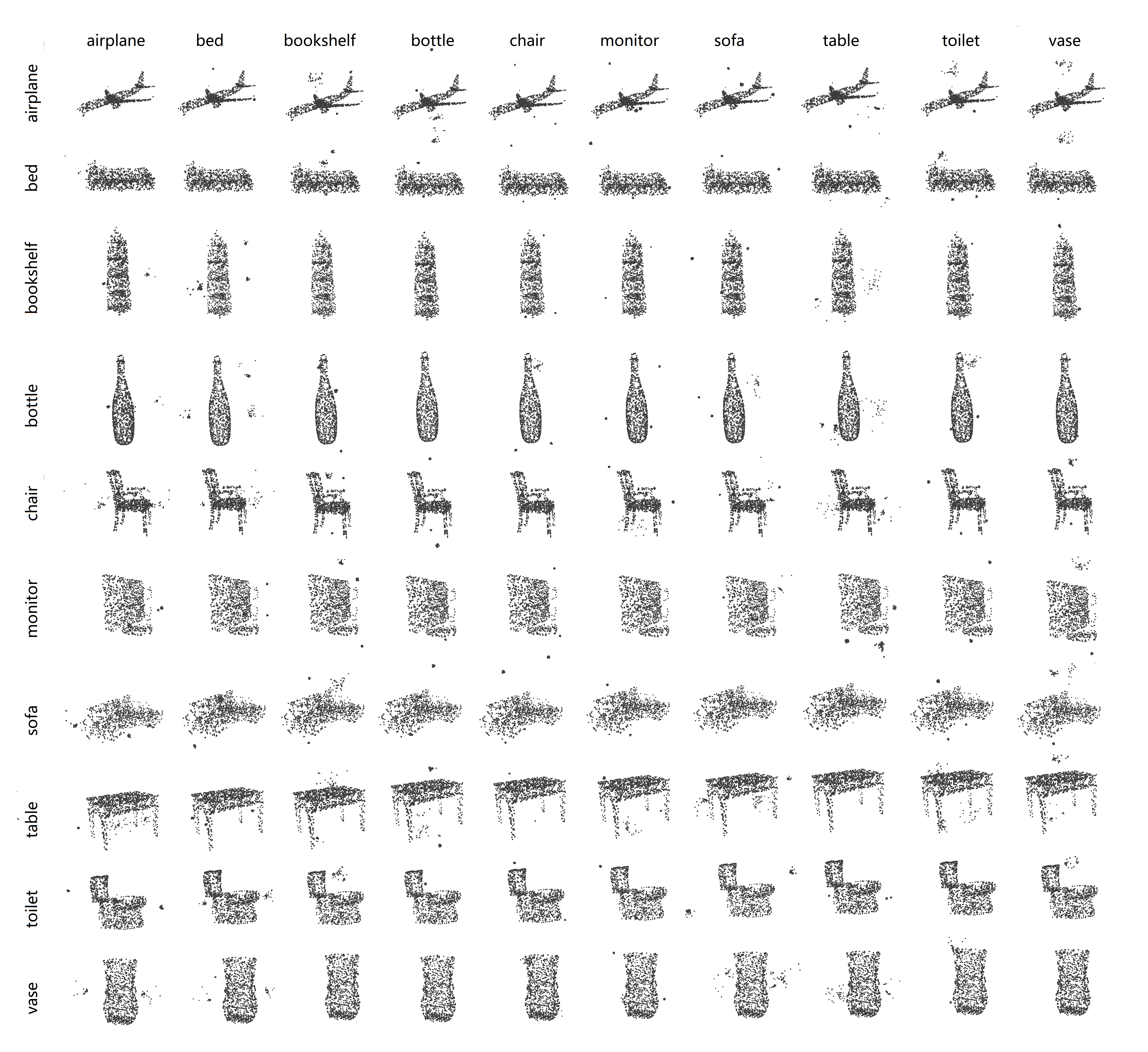}
    \caption{Visualization for adding 3 adversarial clusters (all attack pairs).}
    \label{all_cluster}
\end{figure*}

\begin{figure*}
    \centering
    \includegraphics[width=0.9\textwidth]{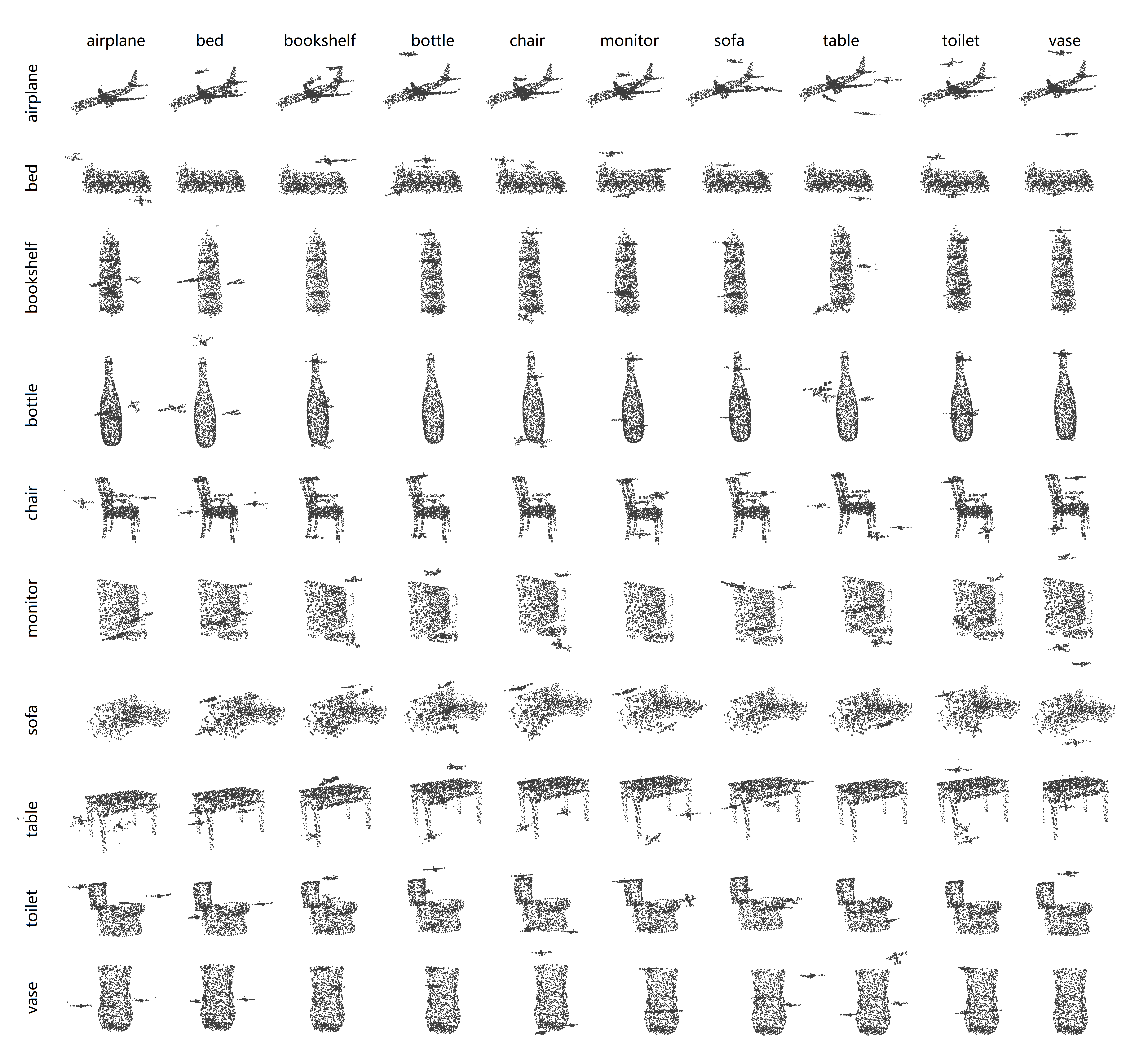}
    \caption{Visualization for adding 3 adversarial objects (all attack pairs).}
    \label{all_object}
\end{figure*}

Besides bottles, here we provide visualizations of victim objects in more categories. The visualization results for \cluster and \object are in Figure~\ref{all_cluster} and Figure~\ref{all_object}, respectively.
% We do not provide additional results for other attacks because they are all indistinguishable. \chong{or should I plot visualization for all five different attacks? Charles: I think it is fine to not present adding and shifting.}

%\begin{figure}[t]
%  \centering
%  \resizebox{1.01\linewidth}{!}{
%  \mbox{
%  \subfloat[]{\includegraphics[width=0.5\linewidth]{img/pert_cdf.png}\label{shift_distribution}}
%   \quad
%  \subfloat[]{\includegraphics[width=0.5\linewidth]{img/add_distribution.png}\label{add_distribution}}
  %}}

  %\caption{Distributions of point perturbation distances. (a) CDF of point shifting distance for \perturbation; (b) Distributions for distance of nearest point pairs in independent point adding attack.}
%\end{figure}

\section{Acknowledgement}
This work is supported by Zhiyuan Scholar Program (Grant No. ZIRC2018-04).

\end{document}